\documentclass[preprint,aps]{revtex4}
\usepackage{graphicx}
\usepackage{mathptmx} 
\usepackage{natbib}

\begin{document}

\title{
Scaling of Tropical-Cyclone 
Dissipation \\
}
\author
{
Albert Oss\'o$^1$,
\'Alvaro Corral$^{2}$  
\&
Josep Enric Llebot$^1$ 
}
\affiliation{
$^1$%
Grup de F\'\i sica Estad\'\i stica,
Facultat de Ci\`encies, 
Universitat Aut\`onoma de Barcelona,
E-08193 Bellaterra, Barcelona, Spain \\
$^2$%
Centre de Recerca Matem\`atica,
Edifici Cc, Campus UAB,
E-08193 Bellaterra, Barcelona, Spain\\
}
\date{\today}



\maketitle
\pagestyle{empty}

{\bf 
The influence of climate variability and global warming
on the occurrence of tropical cyclones (TC)
is a controversial issue 
\cite{Goldenberg_Science,Trenberth,Emanuel_nature05,Landsea_comment,Webster_Science,Chan_comment,Klotzbach,Shepherd_Knutson,Kossin,Elsner08}.
Existing historical databases on the subject are 
not fully reliable
\cite{Kossin,Gray_comment,Landsea_Science06,Landsea_Eos07},
but a more fundamental hindrance is the lack of basic 
understanding regarding the intrinsic nature of tropical 
cyclone genesis and evolution \cite{Emanuel_book}.
It is known that tropical cyclones involve more than a 
passive response to changing external forcing \cite{Emanuel_bams08},
but it is not clear 
which dynamic behaviour best describes them.
%
%
Here we present a new approach based on the
application of the power dissipation index ($PDI$),
which constitutes an estimation of released energy
\cite{Emanuel_nature05},
to individual tropical cyclones.
A robust law emerges for the statistics of $PDI$,
valid in four different ocean basins and
over long time periods. 
In addition to suggesting a 
novel description of the
physics of tropical cyclones in terms of critical phenomena
\cite{Bak_book,Turcotte_book,Christensen_Moloney}, the law allows to quantify
their response to changing climatic conditions,
with an increase in the largest $PDI$ values with sea surface temperature
or the presence of El Ni\~no phenomenon, depending on the basin
under consideration.
In this way, we demonstrate that
the recent upswing in North 
Atlantic hurricane activity 
does not involve TCs
quantitatively different from 
those in
other sustained high-activity periods 
prior to 1970. 
}

One important characterization of a complex phenomenon is given by
the fluctuations in the ``size'' of the phenomenon over successive occurrences
\cite{Bak_book,Turcotte_book,Malamud_hazards,Christensen_Moloney}.
We refer to neither spatial size (area, volume)
nor 
something like the Saffir-Simpson category \cite{Kantha_Eos}; 
rather, we seek 
a physically relevant measure of released energy.
For a tropical cyclone,
a reasonable proxy for this energy has been proposed by 
Emanuel \cite{Emanuel_nature05}, using the $PDI$, defined 
as
$$
PDI \equiv \sum_t v_t^3 \Delta t\ ,
$$
where $t$ denotes time and runs over the entire lifetime of the storm
and
$v_t$ is the maximum sustained surface wind velocity at time $t$. 
In available best-track records, measurements are provided at intervals 
of $\Delta t=6$ hours. 
Note that in this paper the $PDI$ value is associated with an individual 
tropical cyclone, not with the total annual activity in some ocean basin 
\cite{Emanuel_nature05}.

We analyze tropical cyclone best-track records 
for several ocean basins: the North Atlantic and Northeastern Pacific
(from the National Hurricane Center \cite{NOAA}), and
the Northwestern Pacific and the Southern Hemisphere
(from the Joint Typhoon Warning Center \cite{ATCR_report}).
We exclude the North Indian Ocean, due to the small number 
of storms in the most reliable portion of its records.
[For more details, 
see the
Supplementary Information,
including Fig. S2.]

We display in Fig.~1(a) the $PDI$ probability density, $D(PDI)$,
normalized in the usual way
($\int_0^\infty D(PDI) dPDI =1$),
for all four basins. The distributions include all
storms occurring during an extended period, either 1966-2007 or 1986-2007
as indicated in the legend. 
All four distributions (given vertical offsets for clarity)
can be characterized by a power-law decay in their central regions, 
$$
D(PDI) \propto 1/PDI ^\alpha,
$$
where the exponent $\alpha$ is in between 0.95 and 1.25
(Supplementary Information, including Table S1).
Deviations from the power law at small $PDI$ values
can be attributed to the deliberate incompleteness
of the records for ``not significant'' TCs,
whereas the more rapid decay at 
large $PDI$ values is associated with the 
finite size of the basin. That is, the storms with the largest $PDI$ do not
have enough room to last a longer time,
as their tracks are limited by the size of the basin,
which introduces a cutoff in the distribution
(at its separation from the power-law fit, roughly)
(Supplementary Information, Figs. S3-S7). 
Variations in the definition of the $PDI$,
for example excluding times during which the storm 
attains subtropical or extratropical status, do not modify the
shape of the $PDI$ distribution;
the results are also unchanged
when restricted to storms that do not
make continental landfall 
(Supplementary Information, Figs. S8 and S9).

The degree of similarity
between the basins is truly remarkable,
given the variety of formative processes at work.
TCs in the Western Pacific (typhoons)
develop principally from 
the monsoon, for example,
while North Atlantic hurricanes 
are mainly associated with 
easterly waves 
(and the degree of association depends on the intensity of the hurricane) 
\cite{Shepherd_Knutson,Landsea93}.
In addition,
each regional centre or agency follows 
different protocols in obtaining their data,
using techniques which have gradually 
improved \cite{Kossin,Gray_comment,Landsea_Science06,Landsea_Eos07}.

Nevertheless, the shape of the $PDI$ distribution is robust 
over long time periods. Figure~1(b) shows that this consistency holds
over a period of at least 100 years
in the North Atlantic (where the record is longest;
corrections to the calibration of the maximum velocity
do not alter this pattern, Supplementary Information, including Fig. S10).
Thus, even though the database is certainly incomplete 
prior to the satellite era, and even more unreliable 
before aircraft reconnaissance began in 1944,
the fraction of missed storms seems to be independent of 
$PDI$. This finding may appear counterintuitive until we 
consider the fact that a long-lasting tropical cyclone 
might be recorded as two shorter storms if its 
track is lost at some point.
A power-law distribution is robust against
such splitting of the data (Supplementary Information).

The existence of a simple statistical distribution that
describes the whole spectrum of
tropical cyclone sizes 
in different basins over a long period of time 
(apart from 
incompleteness and finite size effects) 
reflects a startling degree of 
unity in the phenomenon;
the small tropical depressions are described in the same
way as the full developed most severe storms. 
Moreover, a robust power-law distribution 
is the hallmark of scale invariance 
\cite{Christensen_Moloney}:
there is no typical tropical cyclone $PDI$,
up to the maximum allowed in a given ocean basin. The fact
that all scales are equally important
for energy dissipation 
poses a great challenge to the modelling
of this complex phenomenon, and even to
large-scale global climate simulations
\cite{Emanuel_bams08}. 

Scale invariance can occur in processes where perturbations 
propagate through a critical (i.e., highly susceptible) medium 
\cite{Bak_book,Turcotte_book,Malamud_hazards}.
Thus, our result could indicate that the atmosphere,
or perhaps the ocean-atmosphere system, is
close to a critical state. 
In fact, this idea is not new;
already in the 70's it was suggested that atmospheric convection takes place 
in a near-unstable environment \cite{Arakawa_schubert}.
Much more recently, Peters and Neelin have demonstrated the
existence of a non-equilibrium stability-instability transition
to which the state of the atmosphere is attracted \cite{Peters_np}.
Some properties of this transition 
can be obtained from 
static images of convecting cloud fields \cite{Peters_percolation} 
or 
local observations of precipitation \cite{Peters_prl}.
These findings support our complex-system approach to 
tropical-cyclone evolution; in correspondence,
we provide a complementary perspective to the puzzle that
these atmospheric processes constitute.

%
%
%
%
%
%

In addition,
an important property of critical systems is that perturbations 
can evolve while keeping a delicate balance between growth and 
attenuation, resulting in sudden intensifications and deintensifications.
Although recent years have seen considerable improvement
in the prediction of tropical cyclone trajectories, 
reliable forecasts of their intensities have not yet been achieved
\cite{Willoughby_eye,verification}.
This failure may not be just due to technical limitations; it may 
be a fundamental feature of the criticality of tropical cyclone evolution.

Tropical cyclone activity shows large interannual variability.
One important factor controlling such variability
is sea surface temperature ($SST$).
We average $SST$ from the Hadley Center \cite{SST}
over the same spatial areas
and months in the TC season than Webster {\it et al.} 
\cite{Webster_Science} in order to get an annual
(so, seasonal) $SST$ value for each basin;
then, we separate
the $PDI$ density into two contributions, 
one for years with $SST$ above its long-term mean value $\langle SST \rangle$ (i.e., high $SST$)
and another one for $SST$ below $\langle SST \rangle$ (low $SST$).
Mathematically,
$\langle SST \rangle \equiv \sum_y SST(y) /Y$,
where $SST(y)$ refers to year $y$
and $Y$ is the total number of years.
%
%
%
In this way
we eliminate the effect of interannual variations in the number of TCs and 
concentrate on a comparison of the individual 
tropical cyclones characterizing each type of year.

Remarkably, for the North Atlantic and the Northeastern Pacific, 
the resulting distributions
have essentially the same shape as the distribution for all years
but with a difference in scale:
high-$SST$ years are characterized by a larger value 
of the finite-size cutoff, 
and conversely for low-$SST$ years, as can be seen in Fig.~2(a).
The other two basins show much minor $PDI$ variation with $SST$.

If we rescale each conditional distribution 
by a power of its mean value,
$\langle PDI \rangle$,
such that
$PDI \rightarrow PDI/\langle PDI \rangle^\nu$ and
$D(PDI) \rightarrow \langle PDI \rangle^\beta  D(PDI)$,
with $\nu=\beta=1$ for $\alpha \le 1$,
and $\nu=1/(2-\alpha)$ and $\beta=\alpha/(2-\alpha)$ for $\alpha > 1$,
it becomes apparent that, for each basin, 
both distributions 
share a similar shape,
as shown in Fig.~2(b).
(The reason of this rescaling is the fact that 
a distribution with $\alpha > 1$ does not scale linearly with its mean value,
see Supplementary Information.)
Then, the difference between high-$SST$ and 
low-$SST$ $PDI$ distributions rests mainly in the 
scale of the finite-size cutoff 
and not in the shape of the distribution.

Years with high $SST$ are thus characterized by hurricanes
with larger $PDI$ values. As the $PDI$ integrates the cube
of the velocity over the storm lifetime,
larger $PDI$ values can result from longer
lifetimes, larger (6-hour) velocities, or both.
An analysis of the distributions of these variables
shows that the increase in $PDI$ comes mainly from 
an increase in the velocities, most apparent
in the range above about 
100 knots (i.e., corresponding to category 3 hurricanes and beyond
\cite{Webster_Science}),
in comparison with years of low $SST$
(Supplementary Information, Fig. S11).

An analogous study can be done as a function of the so-called $MEI$ index
\cite{MEI2},
which quantifies the strength of El Ni\~no phenomenon.
Taking annual values of $MEI$, 
years with $MEI > 0$ (corresponding to El Ni\~no)
lead to increased $PDI$ values in the Northeastern
and Northwestern Pacific, 
but keeping again the same shape of the $PDI$ distribution,
and the opposite for $MEI < 0$,
see Figs. 2(a) and 2(b).
In the case of the Northwestern Pacific,
the contribution of TC durations to the increase
in $PDI$ is larger than in the rest of the cases (Supplementary Information).
%
%
%
This is somehow related to the findings of Refs. \cite{Lander,Emanuel_07}.
Very little variation with $MEI$
is observed in the other two basins. 
Other indices ($AMO$, $NAO$, etc.)
do not seem to influence the $PDI$ 
distribution in any basin.

%

%
%



It is a well-known fact
that individual years of high (or low) TC activity
cluster into longer periods of predominantly high 
(or low) activity. For example, the North Atlantic
has seen extraordinarily high activity between 1995 and 2005, 
which has been linked to global warming through an
increase in the sea surface temperature 
\cite{Emanuel_nature05,Webster_Science}.
The issue is nonetheless controversial 
\cite{Landsea_comment,Chan_comment,Gray_comment}.
According to our analysis, the $PDI$ distribution 
for the period 1995-2005 is indistinguishable 
from the distribution for years with high $SST$
between 1966 and 2007, 
as well as from the distribution
for other periods of high activity, like 1926-1970
\cite{Goldenberg_Science}, see Fig. \ref{highact}.
We conclude that the recent dramatic increase of activity
does not lead to unprecedented energy releases by individual TCs
(although even higher $SST$ could further increase the $PDI$ values).







{\bf Supplementary Information} \\
Available.

{\bf Acknowledgements}\\
We have benefited from the expertise and kindness of
A. Deluca, K. Emanuel,
E. Fukada, A. Gonz\'alez,
J. Kossin, B. Mathiesen, M. Paczuski, O. Peters and A. Turiel.
A.O. and A.C. were put in contact through G. Orriols.
The initial part of our research has been financed by the 
EXPLORA - INGENIO 2010 program, 
and also partially by grants
by other grants of the Spanish MEC and 
Generalitat de Catalunya.

{\bf Author Information}\\
Correspondence and requests for materials should be
addressed to A.C. (ACorral at crm dot cat).

{\bf Competing financial interests}\\
The authors declare no competing financial interests.


\newpage

%
\begin{figure*} 
\caption{
{\bf $|$ Power-law distributions of tropical-cyclone $PDI$ values.}
{ (a)}
$PDI$ probability densities for tropical cyclones in the 
North Atlantic, 
Northeastern Pacific,
Northwestern Pacific, 
and Southern Hemisphere basins. The period considered is either 
1966-2007 or 1986-2007, as indicated in the legend, depending on 
the reliability of the records and the sufficiency of statistics.
The values in the vertical axis are divided by the factors
1, $\sqrt{1000}$, 1000 and $\sqrt{1000^3}$, 
to separate the curves for clarity. From top to bottom, 
the number of tropical cyclones in each case is
469, 674, 690 and 601.
The distributions are consistent with a power law (straight lines)
over most of their range, with exponents
1.19, 1.17, 0.98 and 1.11 ($\pm 0.07$, Supplementary Information).
The Kolmogorov-Smirnov (KS) test yields 
$p-$values larger than 60 \% in all basins (Supplementary Information).
Deviations from the power law at large $PDI$ values reflect 
the finite size effect.
{ (b)} 
$PDI$ probability densities for tropical cyclones
in the North Atlantic over
the 54-year periods 1900-1953 and 1954-2007,
with 436 and 579 storms respectively.
A two-sample KS test (Supplementary Information) gives a $p-$value around $15 \%$ 
($51 \%$ if $PDI > 2\cdot 10^9 \, m^3/s^2$ ).
The robustness of the distribution is apparent despite
a notable lack of homogeneity in the quality of records.
}
\end{figure*}

%
\begin{figure*} 
\caption{
{\bf $|$ Scaling of $PDI$ distributions 
conditioned to the values of $SST$ and $MEI$.}
{ (a)}
$PDI$ probability densities 
calculated separately for years with high $SST$
and for years with low $SST$,
as well as for years with $MEI > 0 $ and for years with $MEI < 0$.
Tropical depressions (storms whose maximum $v_t$ is below
34 knots, 1 knot $=$ 1.85 km/h) are excluded from the Northwestern Pacific
dataset, in order to give all basins the same treatment.
Time periods and
vertical offsets are as in Fig. 1(a).
{ (b)} 
As with the previous panel, except that 
the axes are rescaled by $\langle PDI\rangle^{\nu(\alpha)}$ and 
$1/\langle PDI\rangle^{\beta(\alpha)}$,
with $\alpha=1.19$, 1.17, 1.17 and 1.0
(the data set at the bottom has been shifted extra factors
100 and $10^4$ in each axis
for clarity sake);
the two-sample KS test yields $p-$values 
equal to 90 \%, 73 \%, 44 \% and 73 \%,
for $PDI/\langle PDI \rangle^\nu >$ 0.0003, 0.0005, 0.0005 and 0.05,
in units of $(m^3/s^2)^{1-\nu}$;
the ratios of mean $PDI$ between high and low annual values of $SST$ or $MEI$, 
$\langle PDI \rangle_{high}/\langle PDI \rangle_{low}$, are 
1.4, 
1.6, 
1.4, 
and
1.4, 
always from top to bottom.
}
\end{figure*}

\begin{figure*}
\caption{
{\bf $|$ Equivalence between periods of predominantly high activity
and high $SST$.}
$PDI$ probability densities are compared for the years
1995-2005 and 1926-1970,
as well as for years with high $SST$ during 1966-2007.
The number of storms is
166, 427 and 250, respectively.
Two-sample KS tests yield $p-$values
larger than $40 \%$ for each pair of distributions (Supplementary Information, Table S2);
so we can conclude that no significant difference
in the energy of TCs
can be observed between these periods. 
\label{highact}
}
\end{figure*}

\begin{figure*}
\includegraphics[width=15cm]{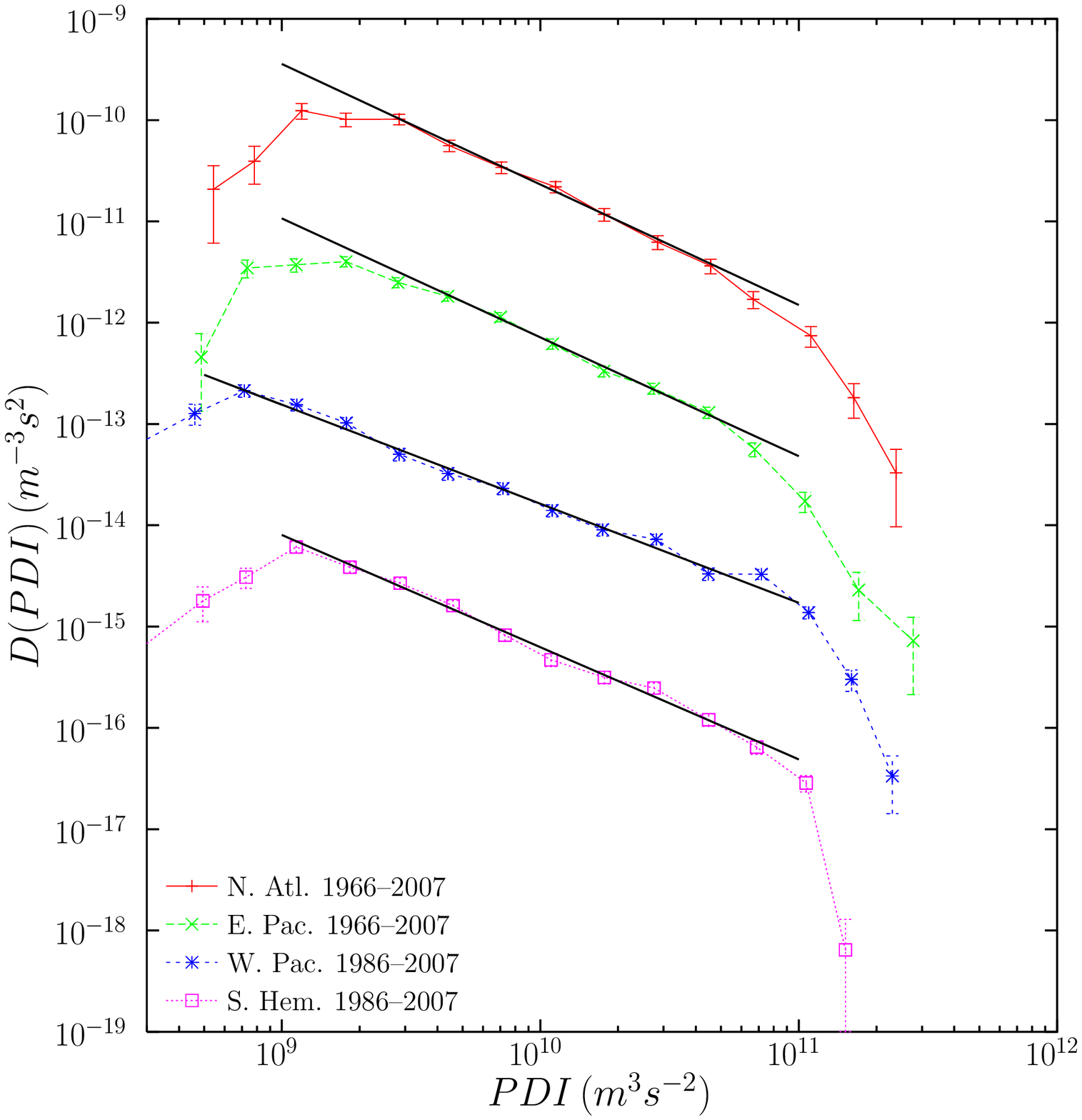}
1(a)
\end{figure*} 
\begin{figure*} 
\includegraphics[width=15cm]{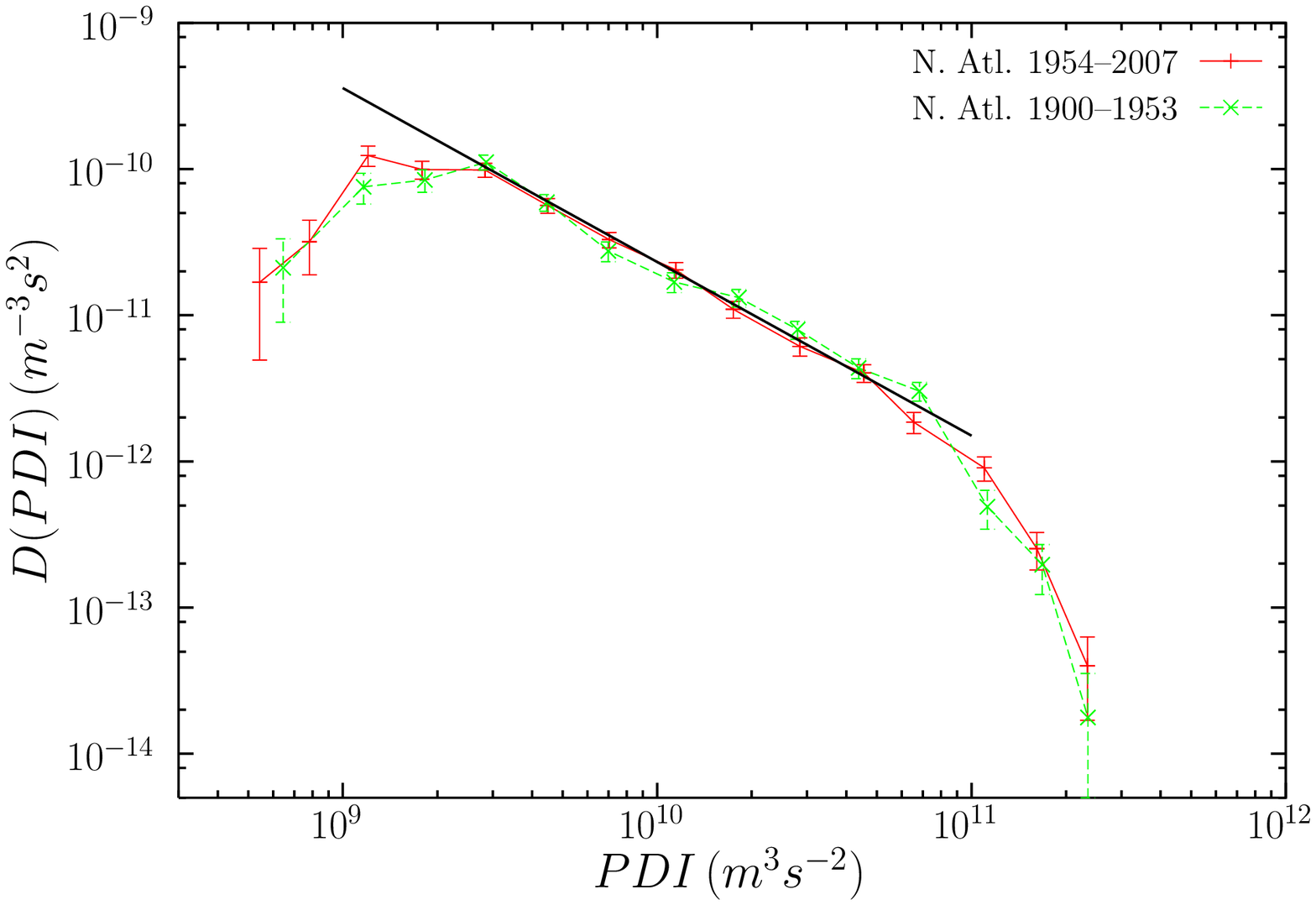}
1(b)
\end{figure*} 

\begin{figure*}
\includegraphics[width=15cm]{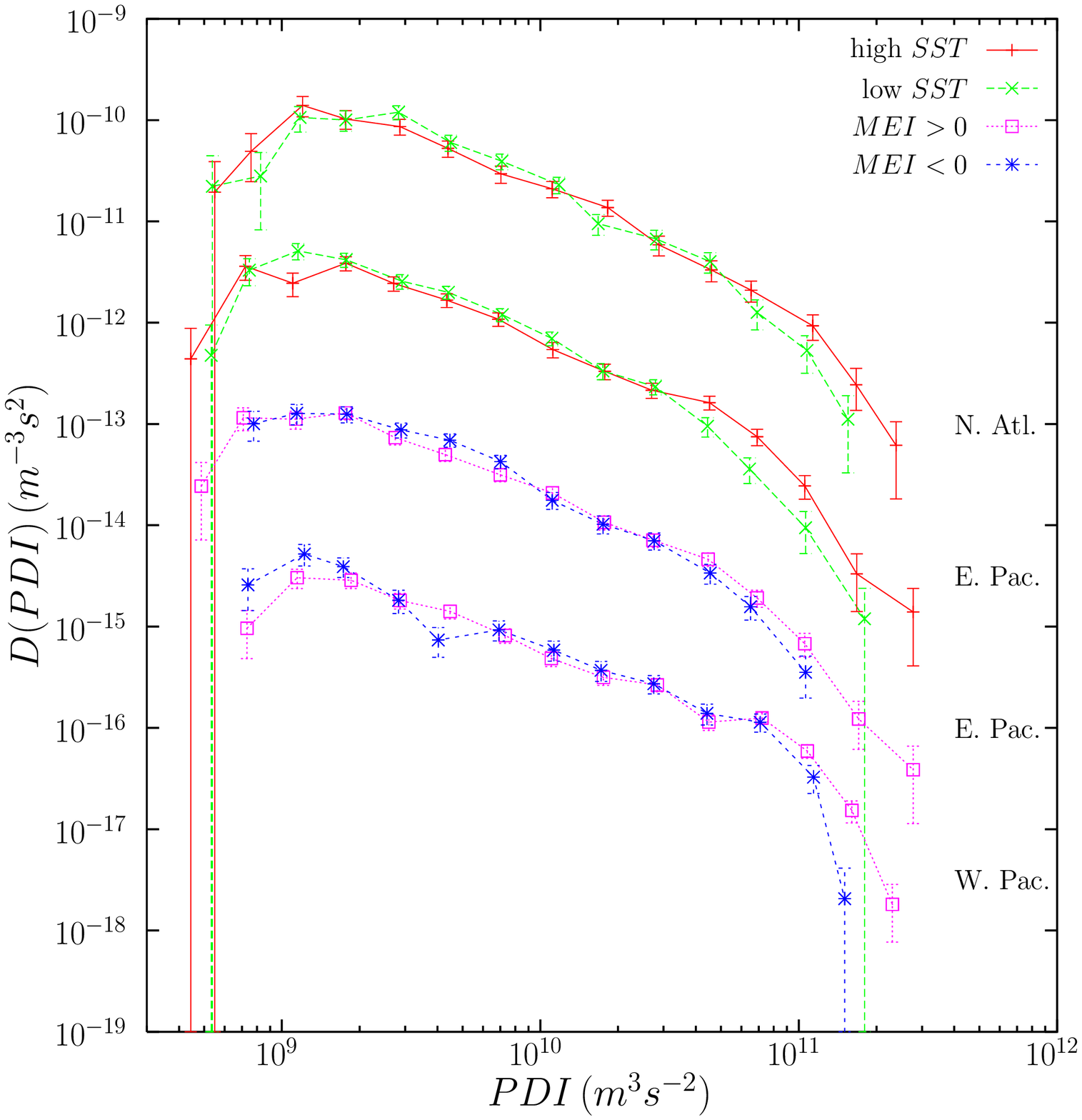}
2(a)
\end{figure*} 
\begin{figure*} 
\includegraphics[width=15cm]{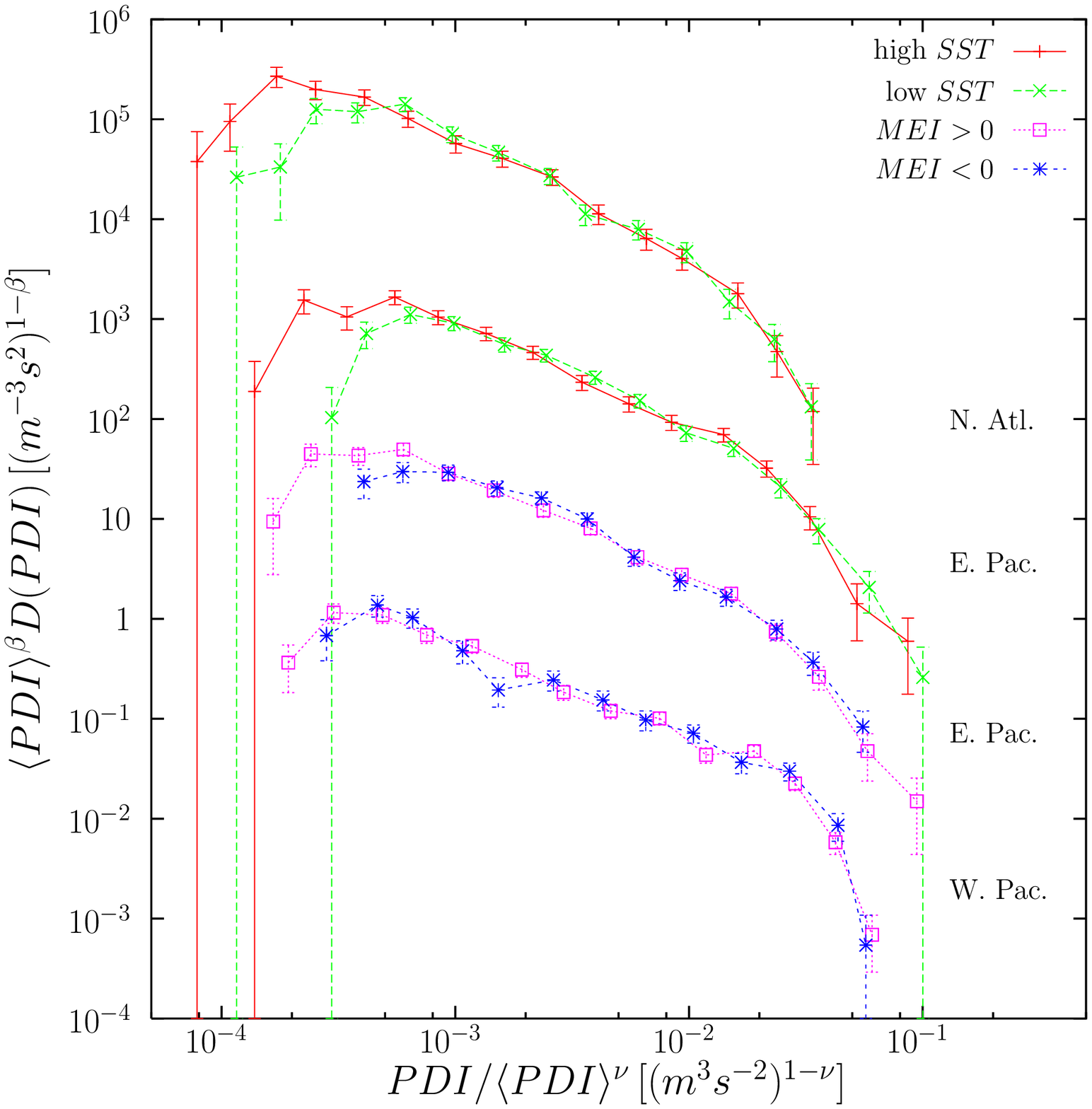}
2(b)
\end{figure*} 

\begin{figure*}
\includegraphics[width=15cm]{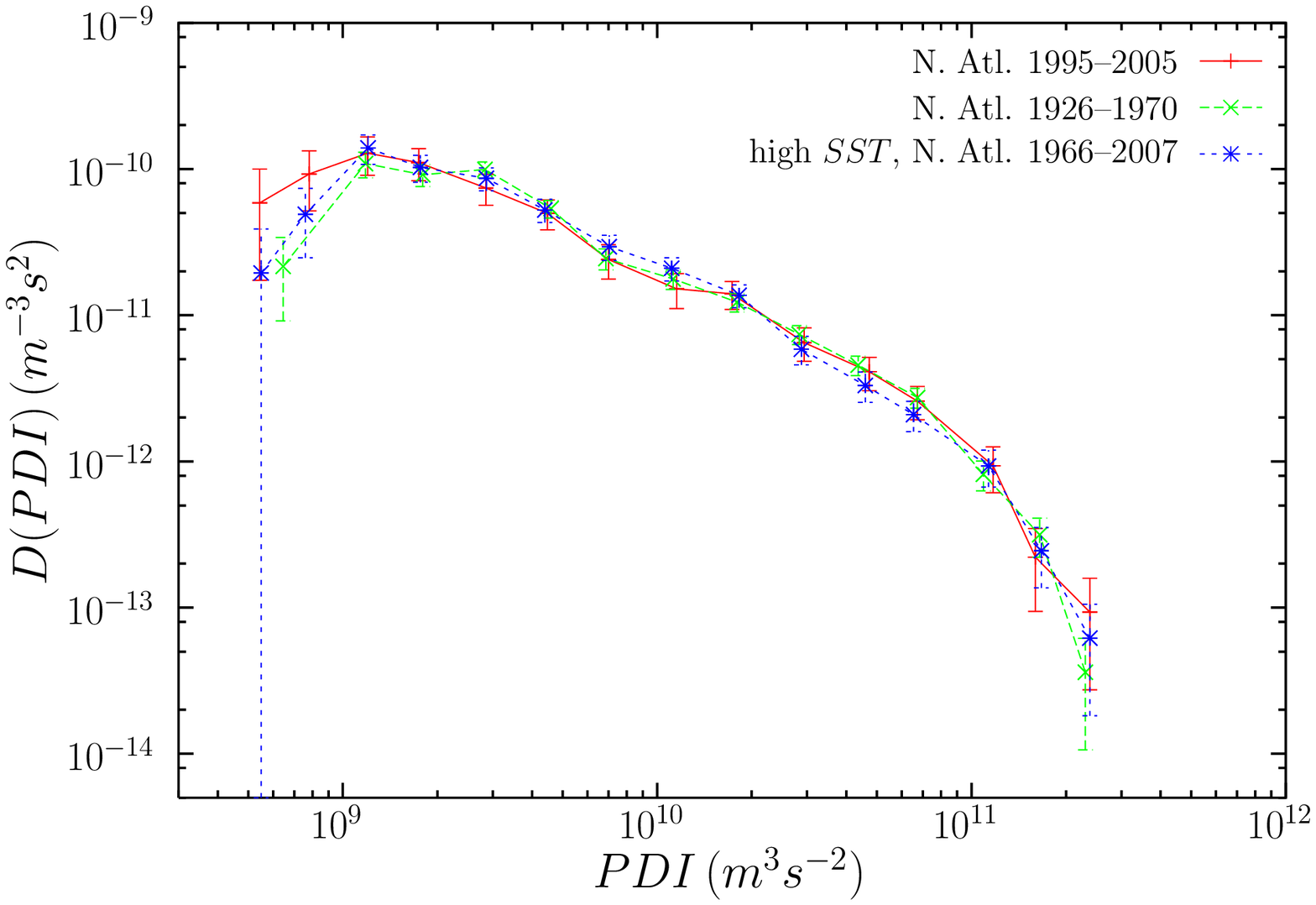}
3
\end{figure*}

\end{document}